\documentclass[aps,10pt,superscriptaddress]{revtex4}
\usepackage{graphicx}
\usepackage{epstopdf}
\usepackage{amssymb}
\usepackage{mathrsfs}
\usepackage{amsmath}
\usepackage{color}
\usepackage{latexsym}
\usepackage{amsfonts}
\usepackage{hyperref}
\usepackage{subfigure}
\usepackage{wasysym}
\usepackage{dcolumn}
\usepackage{bm}
\usepackage[percent]{overpic}
\usepackage{natbib}
\usepackage{booktabs}
\usepackage{gensymb}

\begin{document}

\selectfont
\title{Internal structure and swelling behaviour of {\it in silico} microgel particles}

\author{Lorenzo Rovigatti}
\affiliation{CNR-ISC, Uos Sapienza, Piazzale A. Moro 2, 00185 Roma, Italy}
\affiliation{Dipartimento di Fisica, {\em Sapienza} Universit\`a di Roma, Piazzale A. Moro 2, 00185 Roma, Italy}

\author{Nicoletta Gnan}
\affiliation{CNR-ISC, Uos Sapienza, Piazzale A. Moro 2, 00185 Roma, Italy}
\affiliation{Dipartimento di Fisica, {\em Sapienza} Universit\`a di Roma, Piazzale A. Moro 2, 00185 Roma, Italy}

\author{Emanuela Zaccarelli}\affiliation{CNR-ISC, Uos Sapienza, Piazzale A. Moro 2, 00185 Roma, Italy}
\affiliation{Dipartimento di Fisica, {\em Sapienza} Universit\`a di Roma, Piazzale A. Moro 2, 00185 Roma, Italy}

\definecolor{corr14okt}{rgb}{0,0,1} 
\definecolor{moved}{rgb}{0,0,0}

\begin{abstract} 
Microgels are soft colloids that, in virtue of their polymeric nature, can react to external stimuli such as temperature or pH by changing their size. The resulting swelling/deswelling transition can be exploited in fundamental research as well as for many diverse practical applications, ranging from art restoration to medicine. Such an extraordinary versatility stems from the complex internal structure of the individual microgels, each of which is a crosslinked polymer network. Here we employ a recently-introduced computational method to generate realistic microgel configurations and look at their structural properties, both in real and Fourier space, for several temperatures across the volume phase transition as a function of the crosslinker concentration and of the confining radius employed during the ``in-silico'' synthesis. We find that the chain-length distribution  of the resulting networks can be analytically predicted by a simple theoretical argument. In addition, we find that our results are well-fitted to the fuzzy-sphere model, which correctly reproduces the density profile of the microgels under study.
\end{abstract} 

\maketitle
\section{Introduction}
Microgels are colloid-size particles made by crosslinked polymer networks generally organized into a dense core and a softer, loose corona~\cite{vlassopoulos2014tunable}. They can be synthesized in a range of diameters going from approximately 50 nm to a few microns~\cite{pelton2004unresolved} and their softness can be tuned for example by varying the number of crosslinkers; this allows to generate a plethora of particles with different behaviour ranging from  hard-sphere-like microgels~\cite{schneider2017tuning} for high cross-linkers concentration, to ultrasoft particles when few crosslinks react with monomers~\cite{bachman2015ultrasoft,virtanen2016persulfate}. Besides the tunability in the synthesis protocol, the single most prominent  feature of microgels is the ability to adjust their volume to react to a change of the external conditions~\cite{fernandez2011microgel}; this is the case of the widely exploited  poly-N-isopropylacrylamide (PNIPAM) microgels  which are thermoresponsive and  undergo a volume phase transition (VPT) from a swollen to a collapsed state on increasing temperature~\cite{pelton1986preparation, lyon2012polymer,PhysRevE.94.032601}. Depending on the chemistry and the synthesis protocol, microgels can be designed to be responsive also to pH~\cite{nigro2015dynamic}, salt concentration~\cite{lopez2007macroscopically} or external (\textit{e.g} electric) fields~\cite{crassous2015anisotropic}. Thanks to their inner polymeric architecture, microgel colloids thus possess unique properties which make them remarkably suited for practical applications~\cite{galaev2007smart,fernandez2009gels,oh2008development} as well as for fundamental research\cite{yunker2014physics}.
Indeed, from the latter point view, the possibility of tuning the volume fraction of the samples \emph{in situ} makes it possible to use microgel particles as a model system to explore fundamental physics problems, as highlighted by several recent works~\cite{seth2006elastic,mattsson2009soft,iyer2009self,peng2015two,rossi2015shape}. 
The soft nature of microgels can be incorporated, on a first level, by modelling them as partially-penetrable spheres. Indeed, by leveraging the classic Hertzian approach~\cite{landau1986theory}, which describes the elastic response of a medium in the small-deformation regime, it is possible to derive an effective pair potential that is well-suited to describe microgel suspensions which are not too dense~\cite{zhang2009thermal,pamies2009phase,paloli2013fluid,mohanty2014effective}. However this description neglects the polymeric nature of the microgel, thus failing to describe the behaviour of the suspension at high densities, where strong deformations, interpenetration and entanglement effects play an important role in the interactions among particles~\cite{mohanty2017interpenetration,conley2017jamming}.
In this respect, computer simulations represent a viable tool to test the range of validity, and hence the limits, of the Hertzian approach. Indeed, a numerical model of microgel particles that incorporates the proper degree of detail would allow to probe the realistic effective interactions beyond the simple elastic repulsion.
Attempts in this direction have been made only recently~\cite{Ahuali2017} by leveraging microgels obtained from coarse-grained spherical polymer networks that mimic the swelling behaviour of experimental microgels~\cite{claudio2009comparison,jha2011study,doi:10.1021/la400033s,kobayashi2014structure,ghavami2016internal,Ahuali2017,kobayashi2017polymer}.
Unfortunately, most of these models are based on unrealistic polymer networks, free from entanglements and made by polymer chains of the same length. In fact, in addition to the qualitative agreement with the experimental microgel behaviour, a full understanding of inter-microgel interactions requires  models that are able to account for the inner topology of particles. For instance, non-trivial features of microgels that are deemed to be important are a non-homogeneous distribution of crosslinkers, a continuous distribution of polymer chain-lengths and, arguably, the internal degree of entanglement.

In a recent work~\cite{gnan2017silico}, we have proposed a numerical protocol to design neutral microgel particles based on the self-assembly of gel-forming monomers under spherical confinement.
Once the disordered network is fully assembled, monomers are simulated using classical polymer interactions. With our method we find that the radius of the initial spherical confinement controls to a high degree the internal architecture  of the single particle while maintaining the number of crosslinkers and of monomers constant. In particular, large confinement radii give rise to open and less intertwined polymer networks with larger gyration radii; on the other hand small confinement radii generate small and compact microgel particles. 
Consequently, the different internal structures, obtained at fixed crosslinker concentration and number of monomers,  give rise to different swelling behaviour of the microgels themselves. Using the radius of confinement as an extra parameter of the numerical synthesis we are able to quantitatively reproduce the experimental swelling behaviour of small microgels~\cite{gnan2017silico}.
In this work we exploit the developed method to investigate how the percentage of crosslinkers employed in the numerical synthesis influences the structure and the collapse of the microgel across the VPT. We find that the chain-length distribution across the network, for fixed crosslinker concentration, is not influenced by the confinement but is an intrinsic property of the gel-forming monomers employed for the network assembly. In addition, we demonstrate that the chain-size distribution and the average number of monomers per chain can be predicted with an heuristic argument based on the Flory theory.
Finally, we show that the effect of the crosslinkers is to increase the compactness of microgels with consequences on the swelling behaviour of the resulting particles. We find that, in all the cases investigated, the internal structure of  microgels can be well described in terms of the fuzzy-sphere model, confirming that the assembled particles are made by a compact core surrounded by a lower-density corona, in agreement with experiments~\cite{conley2016superresolution}.

\section{Materials and Methods}

The initial microgel configuration is built by performing simulations of a binary mixture of patchy particles under spherical confinement, as described in Ref.~\cite{gnan2017silico}. The two species of patchy particles can form up to two (species $A$) and four (species $B$) bonds to mimic the behaviour of monomers and crosslinkers, respectively. Bonds are possible only between $A$--$A$ and $A$--$B$ pairs of particles. The microscopic patchy model we employ takes advantage of a recently developed mechanism that greatly enhances equilibration, making it possible to easily generate fully-bonded configurations~\cite{Sciortino2017}. 
We build microgels made by about $41000$ particles with three different values of the crosslinker concentration $c$, namely $c = 1.4\%$, $3.2\%$ and $5.0\%$, and several values of the radius of confinement $Z$, ranging from $Z = 30\sigma$ to $Z = 70\sigma$ where $\sigma$ is the monomer size. 

Once the initial configurations are generated, the confinement is removed and the resulting topology is frozen-in by replacing the patchy force field with the Kremer-Grest set of interactions~\cite{kremer1990dynamics}, which models polymers in a good solvent. Within this framework, the covalent bonds are modelled through a finite extensible non-linear elastic (FENE) term as given by

\begin{eqnarray}
\label{eq:V_FENE}
V_{\rm FENE}(|\vec r_{ij}|)=
\begin{cases}
-\varepsilon \, k_{F}R_0^2 \ln(1-(\frac{r_{ij}}{R_0\sigma})^2) & {\rm if} \quad |\vec r_{ij}|< R_0\sigma \\[0.5em]
0 & {\rm otherwise.}
\end{cases}
\end{eqnarray}
\noindent
where $\sigma$ is taken as the unit of length, $\epsilon$ controls the energy scale, $R_{0}=1.5$ sets the maximum extension of the bond and $k_F=15$ is the spring constant.

The generic monomer-monomer steric repulsion is given by a Weeks-Chandler-Andersen~\cite{wca} term which reads

\begin{eqnarray}
V_{\rm WCA}(r) = 
\begin{cases}
4\epsilon\left[\left(\frac{\sigma}{r}\right)^{12} - \left(\frac{\sigma}{r}\right)^6 \right] + \epsilon & {\rm if} \quad r \leq 2^\frac{1}{6}\sigma \\[0.5em]
0 & {\rm if} \quad r > 2^\frac{1}{6}\sigma
\end{cases}
\label{eq:WCA}
\end{eqnarray}

The possibility of tuning the quality of the solvent is provided by an additional interaction term, $V_{\alpha}$, acting between all monomer pairs. The strength of this term is controlled by the parameter $\alpha$, which implicitly sets the solvophobicity of the monomers and plays the role of an inverse temperature~\cite{soddemann2001generic,verso2015simulation}. This attractive term reads,
\begin{eqnarray}
\label{eq:V_alpha}
V_{\alpha}(r) =
\begin{cases}
-\varepsilon\alpha & {\rm if} \quad r \leq 2^{1/6} \sigma\\[0.5em]
\frac{1}{2}\alpha\varepsilon [\cos(\gamma (r/\sigma)^2+\beta) -1] & {\rm if} \quad 2^{1/6}\sigma < r\leq R_{0} \sigma\\[0.5em]
0 & {\rm otherwise}
\end{cases}
\end{eqnarray}
\noindent where $\gamma=\pi (2.25-2^{1/3})^{-1}$ and $\beta=2\pi-2.25\gamma$~\cite{soddemann2001generic}.

We simulate single microgels by means of molecular dynamics simulations performed in the isothermal ensemble at constant reduced temperature, $T^*=k_BT/\epsilon=1.0$ where $k_B$ is the Boltzmann constant and $k_B/\epsilon=1$. The temperature is kept fixed by a Nos\`{e}-Hoover thermostat and the integration is carried out with a leap-frog scheme with a reduced time step $\delta t^*=\delta t\sqrt{\epsilon/ (m  \sigma^2 ) }=0.001$, where $m$ is the mass of the monomer which is set to $m=1$. We vary the solvophobic parameter from $\alpha=0$ (where the system is in a good solvent) to $\alpha=1.5$, for which a complete collapse of the microgel is observed. We average all the investigated quantities over a number of distinct initial configurations, depending on the radius of the initial confinement.

\section{Results}

\begin{figure}[h!]
\includegraphics[width=0.5\textwidth,clip]{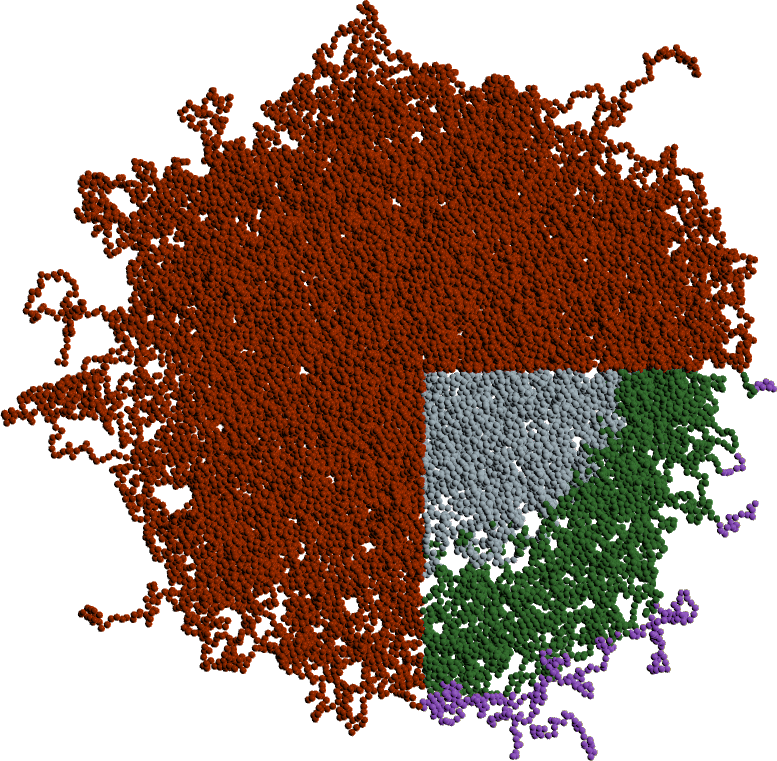}
\caption{\label{fig:sketch} A simulation snapshot of a microgel generated with $c=1.4\%$ and $Z=50\sigma$. The bottom right-hand corner shows a slab of thickness $20 \sigma$, cut out to showcase the radial heterogeneity of microgels. In this part, monomers are coloured according to their distance $d$ from the microgel centre of mass: for this specific case, grey monomers with $d \lesssim 34\sigma$ belong to the core region, green monomers with $34 \sigma \lesssim d \lesssim 52 \sigma$ are part of the corona, while violet monomers with $d\lesssim 52\sigma$ compose the so-called dangling chains.}
\end{figure}

Figure~\ref{fig:sketch} provides a visual representation of a portion of a microgel generated \textit{in silico} with the method described above. 
The disordered nature of the polymer network is evident. Monomers are coloured according to their distance $d$ from the microgel centre of mass to illustrate the presence of (at least) three different regions inside the microgel structure. In the centre there is a uniformly dense core, followed by a corona, whose average density decreases as the distance from the centre of mass increases. Finally, there is an extremely diluted outermost layer composed of particles that are part of long chains or loops that stick out of the corona. These \textit{dangling ends} are usually not accounted for by the models employed to describe microgels, such as for example the widely used fuzzy sphere model described below, but they are important for determining a number of properties of the microgels, such as their hydrodynamic radius $R_H$ or their effective interactions~\cite{boon2017swelling,gnan2017silico}.

In what follows we will quantitatively analyse the composition of the polymer network, its shape and size and how these are affected by a change in the quality of the solvent.

\subsection{Number of chains, average chain-length and size distribution}

The behaviour of a polymer network is highly sensitive to its topology~\cite{rubinstein2003polymer}. We therefore start the analysis of the structure of single microgels by looking at the properties of the chains that make up the polymer networks. To this aim we take advantage of the self-assembly process of the employed binary mixture of patchy particles to build the initial network. Following Flory~\cite{flory1953principles}, in a binary mixture of $N_A$ bifunctional particles and $N_B$ particles with functionality $f_B>2$, all having identical reactive sites, the distribution $N_l$ of the size $l$ of the chains that connect the branching points can be written as
\begin{equation}
\label{eq:csd_flory}
N^{\rm Flory}_l = N_A (1 - p_A p_b)^2 (p_A p_b)^{l - 1}
\end{equation}
where $p_A = \frac{2 N_A}{2 N_A + f_B N_B}$ is the fraction of patches of type $A$ and $p_b$ is the bonding probability (in the language of Flory, $p_b$ is the probability that a randomly-chosen site has reacted). $N_l^{\rm Flory}$ is normalised such that $\sum_{l=1}^\infty l N_l = N_A$ and $\sum_{l=1}^\infty N_l = N_c$, where $N_c$ is the total number of chains. Consequently, the average chain length in the Flory approach $\langle l \rangle^{\rm Flory}$ is given by
\begin{equation}
\langle l \rangle^{\rm Flory} = \frac{\sum_{l=1}^\infty l N_l}{\sum_{l=1}^\infty N_l} = \frac{N_A}{N_c}.
\end{equation}
\noindent
In the fully-bonded limit, $p_b \to 1$ and hence $N_l^{\rm Flory} \to N_A (1 - p_A)^2 (p_A)^{l - 1}$, $N_c \to N_A (1 - p_A) = N_A p_B$ and $\langle l \rangle^{\rm Flory} \to p_B^{-1}$, where $p_B = 1 - p_A$ is the fraction of patches of type $B$~\cite{bianchi_FS}. The same result can be obtained by considering that, for $p_b \to 1$, the number of $B$-sites connected to $A$-sites, which is, by definition, also the number of chain ends, is $N_e = f_B N_B p_A = 2 N_A p_B$. Since each chain has two ends, $\langle l \rangle^{\rm Flory}  \to N_A/N_c = 2 N_A/N_e = p_B^{-1}$.

In the system considered here, bonding between $B$ sites is forbidden and hence the Flory approach does not hold. However, we can still compute the total number of chain ends in a fully-bonded system by using a similar procedure. Indeed, since $B$-sites can only be bonded to $A$-sites, then $N_e = f_B N_B$. It follows that, for $p_b \to 1$,
\begin{equation}
\label{eq:asympt_avg_length}
\langle l \rangle^{\rm asympt}= \frac{2N_A}{f_B N_B} = \frac{p_A}{p_B}.
\end{equation}
\noindent
Since $\langle l \rangle^{\rm asympt}= \sum_{l=1}^\infty l N_l / \sum_{l=1}^\infty N_l$ and $\sum_{l=1}^\infty l N_l = N_A$, if we assume that, following Eq.~\eqref{eq:csd_flory}, the chain-size distribution has an exponential form, we obtain

\begin{equation}
\label{eq:asympt_csd}
N_{l}^{\rm asympt}= N_A \left( \frac{p_B}{p_A} \right)^2 \left( \frac{p_A - p_B}{p_A} \right)^{l-1}
\end{equation}

\begin{figure}[h!]
\includegraphics[width=0.5\textwidth,clip]{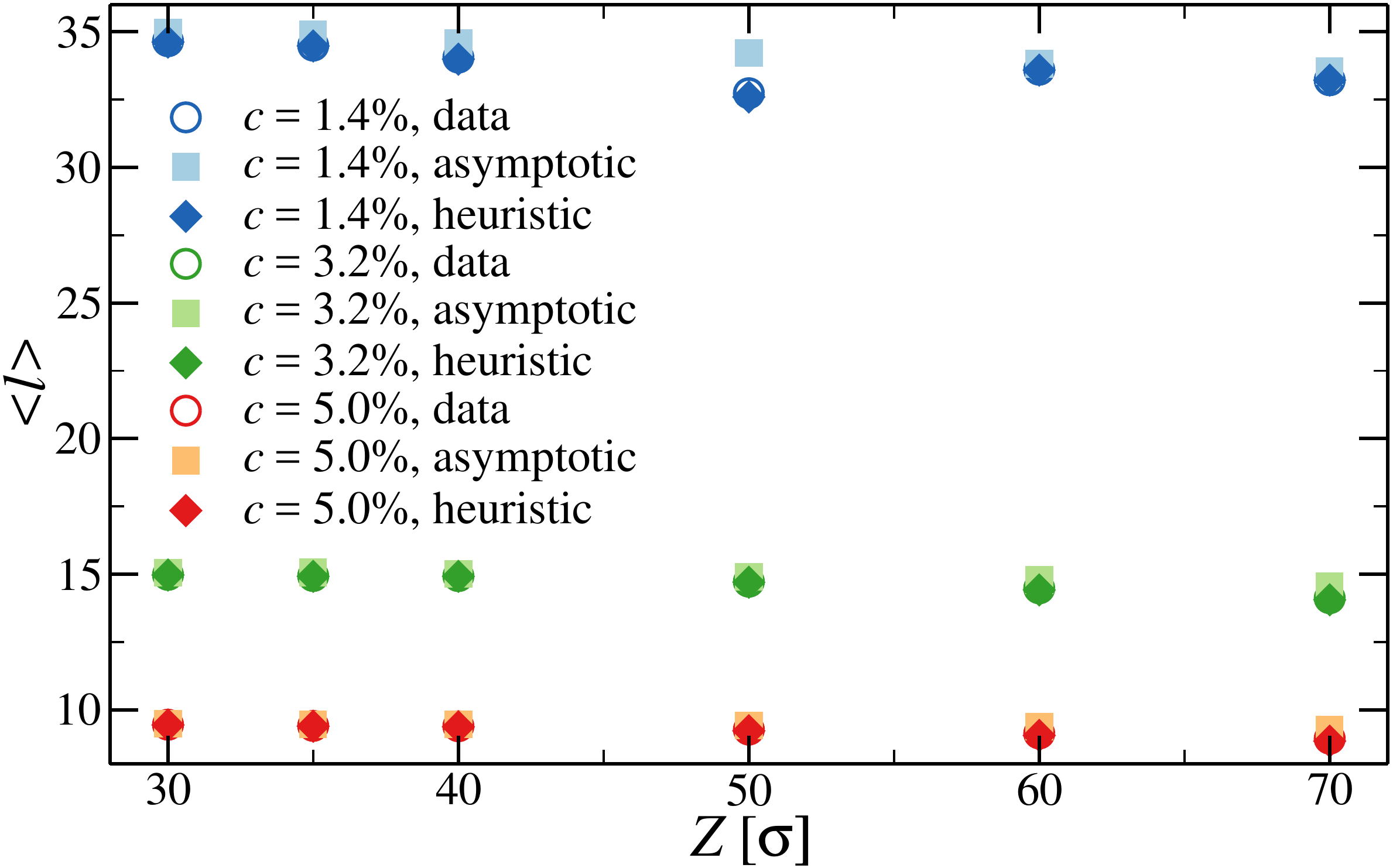}
\caption{\label{fig:avg_length}Average chain length $\langle l \rangle$ for microgels with different crosslinker concentrations as a function of the initial radius of confinement. Open circles are simulation data, filled light squares and dark diamonds are the asymptotic (Eq.~\eqref{eq:asympt_csd})
and non-asymptotic (Eq.~\eqref{eq:phenomenological_avg_length}) theoretical values, respectively. Simulation data are always almost completely hidden behind the results of the heuristic theory, showcasing the very good agreement between the two.}
\end{figure}

Figure~\ref{fig:avg_length} shows the average chain length $\langle l \rangle$ for all the generated microgels comparing the numerical results with the theoretical ones. In particular,  the asymptotic ($p_b \to 1$) values, computed according to Eq.~\eqref{eq:asympt_avg_length},
are always a few percent larger than the simulation data.
The agreement between theory and data, which is already very good, can be further improved by reintroducing the dependence of $N_l$ on $p_b$, which is always very close to but not exactly one ($p_b \gtrsim 0.998$). This operation can be heuristically performed by noting that, in the original Flory expression, Eq.~\ref{eq:csd_flory}, taking the $p_b \to 1$ limit is effectively equivalent to substituting $p_A p_b$ with $p_A$. By making the opposite substitution in Eq.~\ref
{eq:asympt_csd} we obtain the following $p_b$-dependent expressions for $N_l$ and $\langle l \rangle$ for our model:
\begin{align}
\label{eq:phenomenological_csd}
N_l & = N_2 \left( \frac{1 - p_Ap_b}{p_Ap_b} \right)^2 \left( \frac{2p_Ap_b - 1}{p_Ap_b} \right)^{l-1}\\
\label{eq:phenomenological_avg_length}
\langle l \rangle & = \frac{p_Ap_b}{1 - p_Ap_b}.
\end{align}
\noindent
We stress that the above relations were not formally derived and therefore should be considered of heuristic nature. Nonetheless, we find a nearly perfect overlap between simulation data and values calculated from Eqs~(\ref{eq:phenomenological_csd})-(\ref{eq:phenomenological_avg_length}), as shown in Fig~.\ref{fig:avg_length}. 

\begin{figure}[h!]
\includegraphics[width=0.5\textwidth,clip]{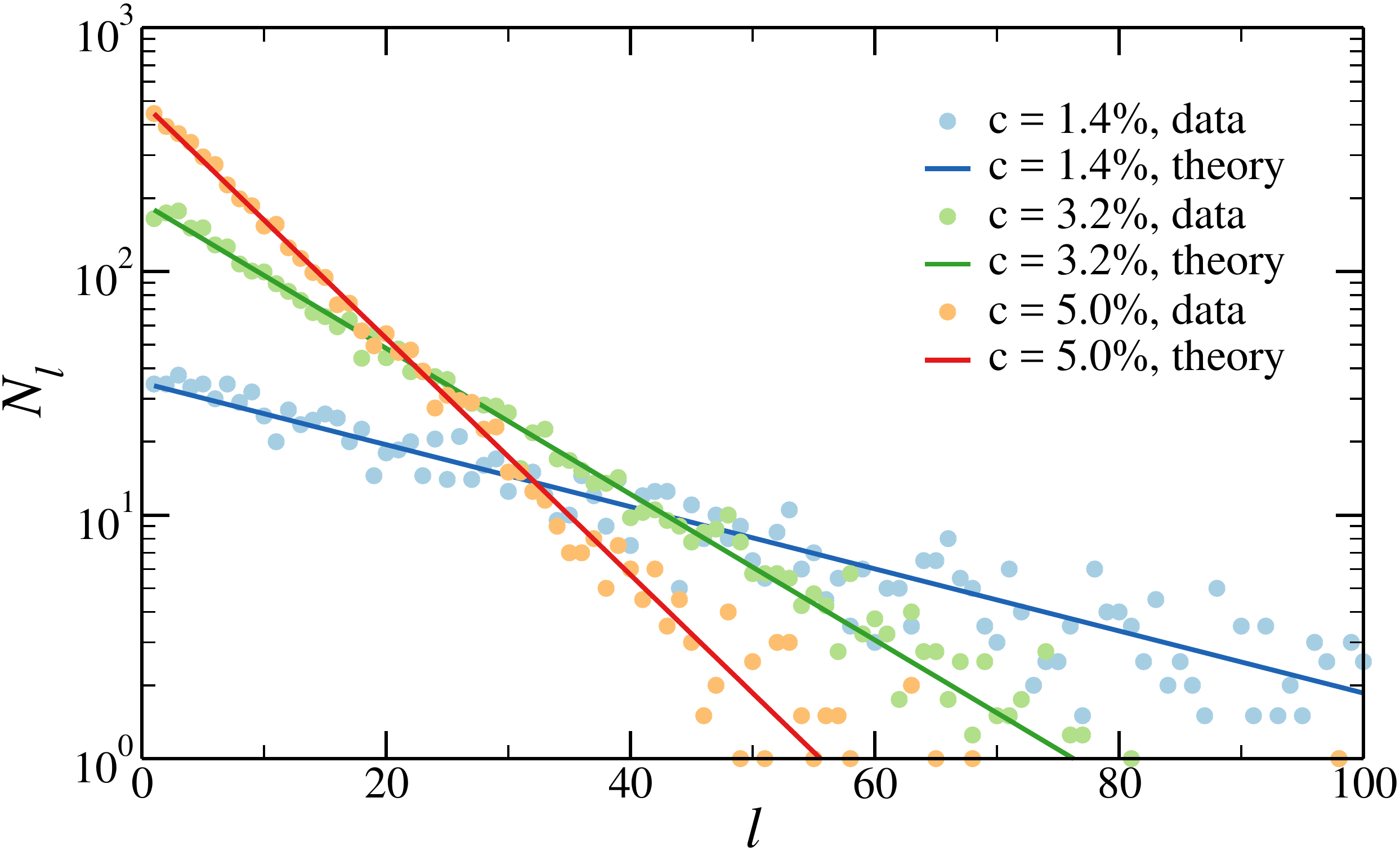}
\caption{\label{fig:csd_R40}Simulation (points) and theoretical (lines) chain-size distributions $N_l$ for microgels with different crosslink concentrations generated with an initial radius of confinement $Z = 40\sigma$. The theoretical curves come from Eq.~\ref{eq:asympt_csd}.}
\end{figure}

We can go one step further and compare the theoretical and numerical chain-size distributions. Fig.~\ref{fig:csd_R40} shows $N_l$ for microgels with different crosslink concentrations and $Z = 40\sigma$, comparing the simulation data (points) with the theoretical expression given by Eq.~\eqref{eq:asympt_csd}. We note that the use of Eq.~\eqref{eq:asympt_csd} or Eq.~\eqref{eq:phenomenological_csd} to calculate $N_l$ yields indistinguishable results within the statistical noise of the data. We thus find that the chain-size distributions are always exponential, supporting the inital assumption used in the derivation of the theoretical expression. Moreover, the agreement between simulation data and theory is excellent in the whole range of sizes observed in simulation. We note on passing that the same qualitative picture (exponential $N_l$, excellent agreement with Eq.~\eqref{eq:asympt_csd}) holds true for all the investigated microgels.

These results confirm that the properties of the chains do not depend on the choice of the confinement radius $Z$. Thus, this is a useful parameter which allows to control the size and the swelling behavior of the microgels, but it does not affect the connective properties of the chains. However, the network topology and thus the number of entanglements can be varied in this way, and we will consider their characterisation in future work. 

\subsection{Form factors and density profiles in the swollen regime}

Experimentally, the structure of microgels is most often probed with neutron or X-ray scattering techniques. The output of such experiments is the form factor $P(q)$, which provides a description of the microscopic structure in reciprocal space. The form factor itself is then usually fitted to a model with a minimal amount of adjustable parameters to obtain the structure in real space. In the case of microgels, the most used model is the so-called \textit{fuzzy sphere} model~\cite{stieger2004small}, which assumes a constant-density core of radius $R'$ and a smearing parameter $\sigma_{\rm surf}$ that is linked to the size of the outer corona. The static and dynamic inhomogeneities of the underlying polymer networks are accounted for by a Lorentzian term of amplitude $I(0)$ and correlation length $\xi$. All in all, the fuzzy-sphere model gives the expression
\begin{equation}\label{eq:fuzzy}
P(q)=\left[\frac{3 [ \sin(qR') -qR' \cos(qR')] }{(qR')^3} \exp\left(\frac{-(q\sigma_{\rm surf})^2}{2}\right)\right]^2 + \frac{I(0)}{1+\xi^2 q^2}.
\end{equation}

\begin{figure}[h!]
\includegraphics[width=0.5\textwidth,clip]{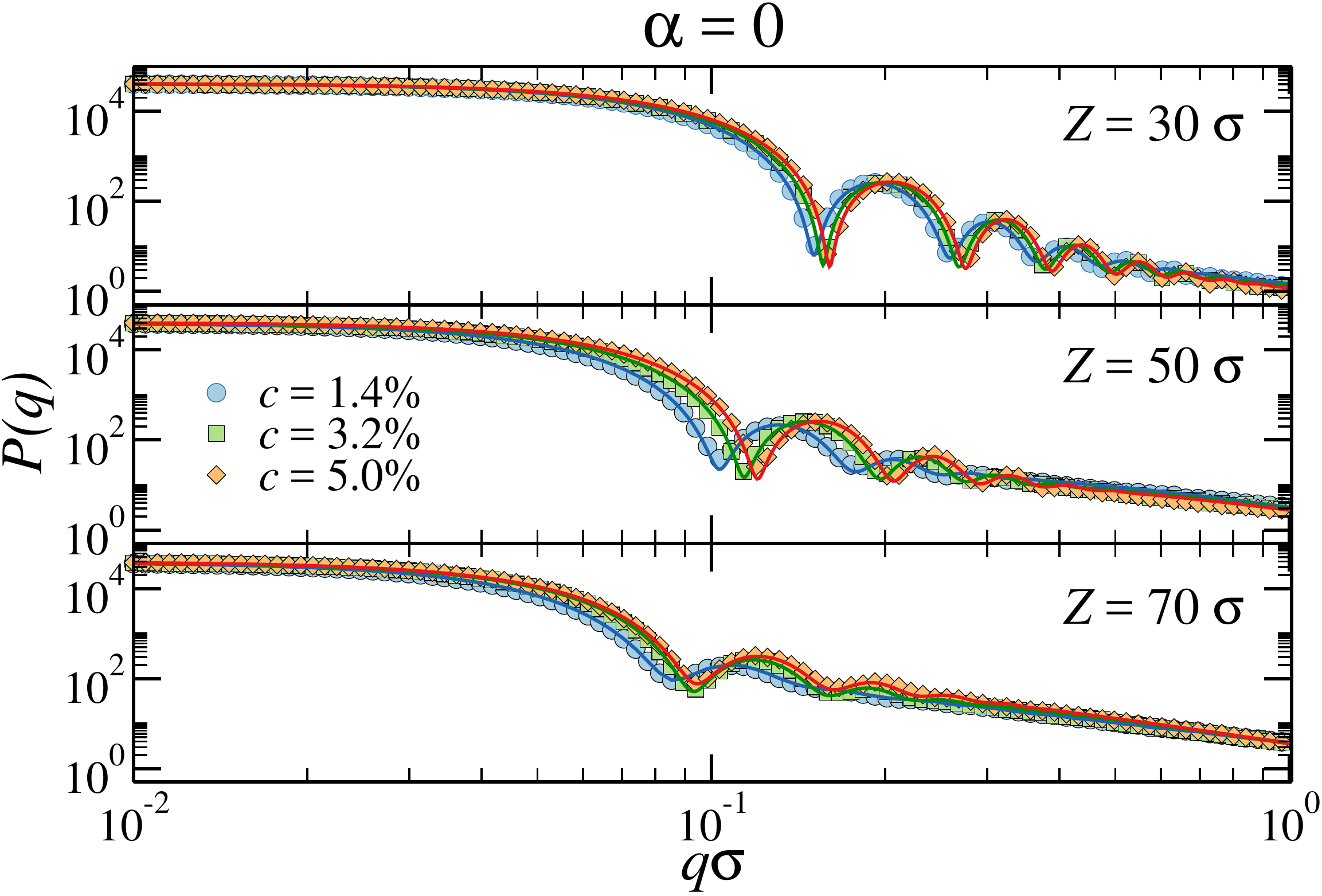}
\caption{\label{fig:pq_alpha0} Form factors  $P(q)$ directly calculated from simulations (points) and obtained from a fuzzy-sphere model (Eq.~\eqref{eq:fuzzy}) fit (lines) for microgels of different crosslinker concentrations at three different values of the radius of confinement $Z$.}
\end{figure}

\begin{figure}[h!]
\includegraphics[width=0.55\textwidth,clip]{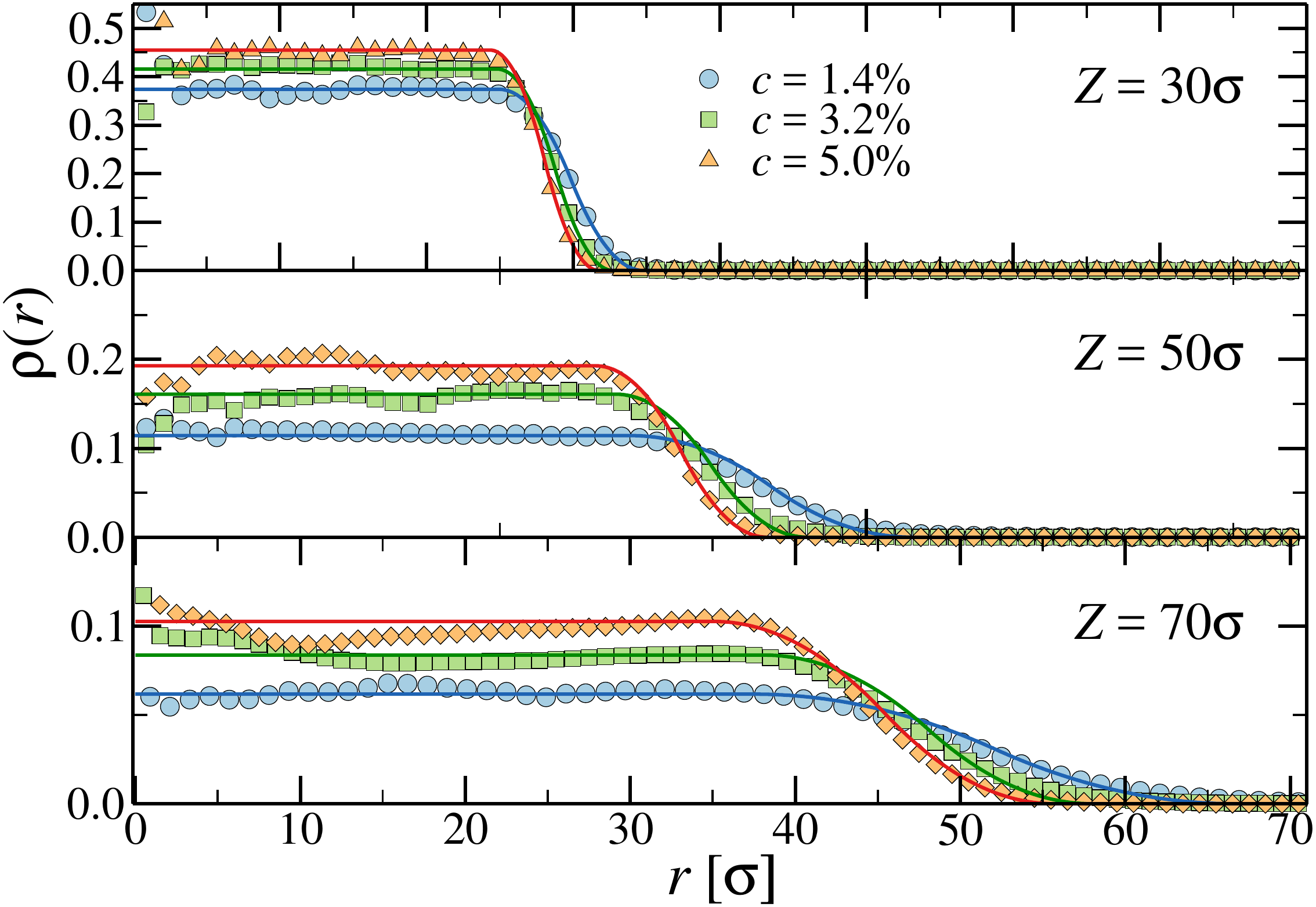}
\caption{\label{fig:density_profile_alpha0}Simulation (points) and fitted (lines) density profiles $\rho(r)$ for microgels of different crosslinker concentrations at two different values of the radius of confinement $Z$.}
\end{figure}

In numerical simulations, the form factor can be directly calculated as $P(q)=(1/N)\sum_{ij}\langle\exp(\vec{q}\cdot\vec{r}_{ij})\rangle$, where the sum runs over all the monomer couples $i,j$ and the angular brackets represent an average taken on an ensemble of configurations. Figure~\ref{fig:pq_alpha0} shows the form factors of numerical microgels at all the investigated crosslinker concentrations and for three values of the radius of confinement. The $P(q)$ is averaged from two to four different realizations of microgels for each studied case.  At a first glance we already see that the position and number of the peaks, which are linked to the size and softness of the microgel, respectively, depend on both $c$ and $Z$. The effect of varying either of these parameters is qualitatively similar: microgels generated with higher $c$ or smaller $Z$ are compact in size and more structured and \textit{vice versa}.
The effect of varying $c$ and $Z$ can be quantified by fitting the numerical $P(q)$ to Eq.~\eqref{eq:fuzzy}. The fit allows to extract the parameters $R'$ and $\sigma_{\rm surf}$, which are needed in order to predict the density profile of the microgel, which reads~\cite{stieger2004small}:
\begin{eqnarray}
\label{eq:profiles}
\frac{\rho(r)}{\rho_0}=
\begin{cases}
1 & {\rm if} \quad r < R_c \\[0.5em]
1 - \frac{(r - R' + 2\sigma_{\rm surf})^2}{8\sigma_{\rm surf}^2} & {\rm if} \quad R_c\leq r < R' \\[0.5em]
\frac{(R' - r + 2\sigma_{\rm surf})^2}{8\sigma_{\rm surf}^2} & {\rm if} \quad  R'\leq r< R'' \\[0.5em]
0  &{\rm if} \quad   r\geq R''
\end{cases}
\end{eqnarray}
\noindent where $R_c=R'-2\sigma_{\rm surf}$ is the radius of the constant-density part of the particle, $\rho_0$ is its number density and  $R''=R'+2\sigma_{\rm surf}$ is the total radius, including the fuzzy shell. The solid lines in Figure~\ref{fig:density_profile_alpha0} are obtained from Eq.~\eqref{eq:profiles} using the parameters extracted from the form factors. The agreement is excellent for all the microgels in the whole investigated $q$-range. The oscillatory character of the density profile in the core region for $Z=70\sigma$ marks the importance of averaging over several realizations, particularly for the case of large confinement radii. In all cases,  a core-corona structure is retained with a sharper variation of the profiles when the confining radius is smaller and with increasing crosslinker concentration.

\subsection{Swelling behaviour}

The ability to respond to a change of the external conditions is the most important property of microgels. The response to external stimuli is finely dependent on a large number of parameters, from the chemical composition of the particle to the synthesis protocol as well as the physical nature of the conditions that are being changed. Here we investigate how \textit{in-silico} microgels generated with different crosslinker concentrations and radius of confinement change their size when the quality of the solvent varies. This is achieved in our model by the variation of the control parameter $\alpha$.
\begin{figure}[h!]
\includegraphics[width=0.5\textwidth,clip]{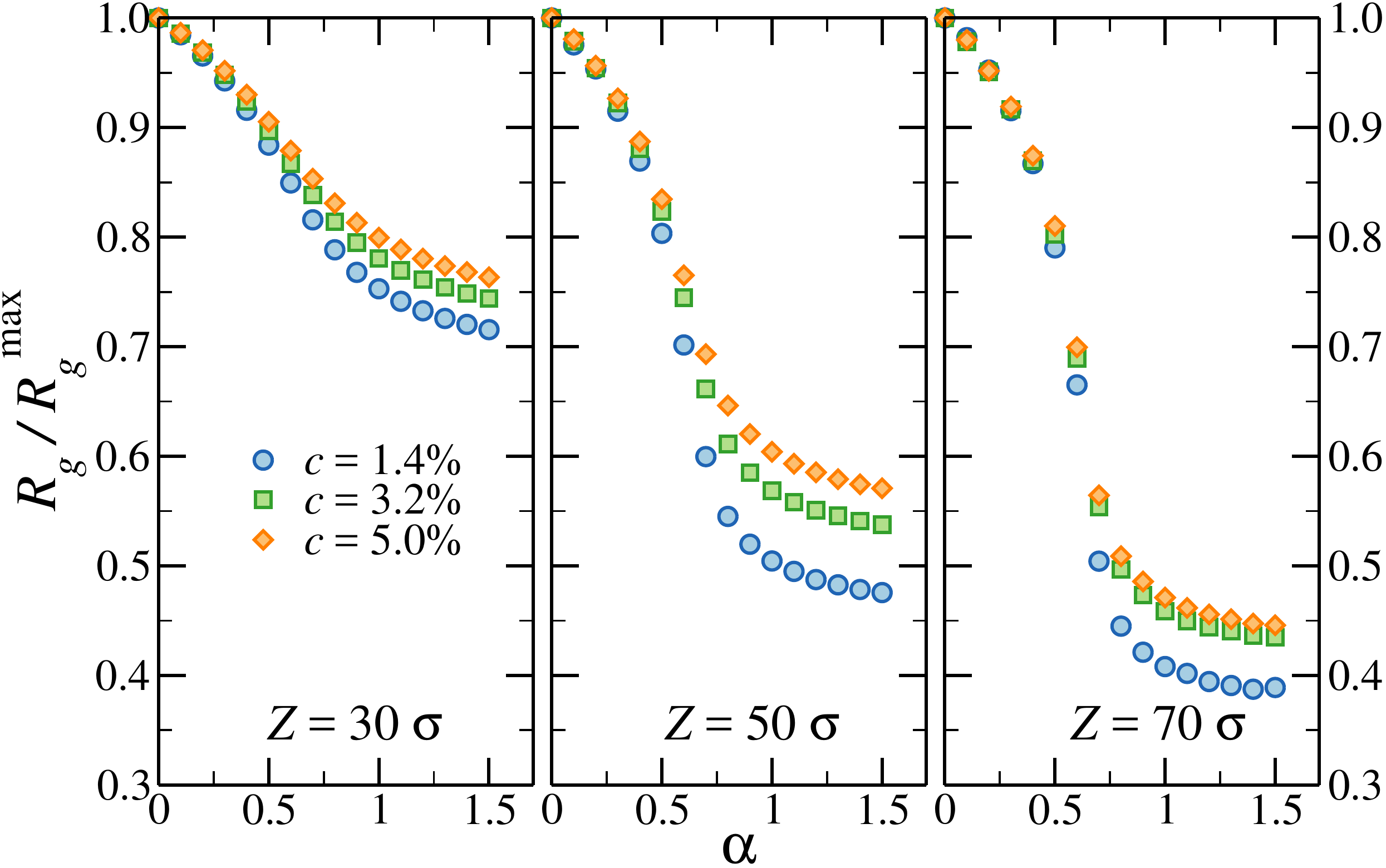}
\caption{\label{fig:swelling}Swelling curves of microgels generated with different crosslinker concentrations and $Z = $ (left) $30\,\sigma$, (middle) $50\,\sigma$ and (right) $70\,\sigma$.}
\end{figure}

Figure~\ref{fig:swelling} shows the so-called swelling curves for all investigated values of $c$ and for three selected values of $Z$ as a function of $\alpha$. These plots show the gyration radius $R_g$ of the particles, normalised by dividing by $R_g^{\rm max} \equiv R_g(\alpha = 0)$ for the maximally swollen conditions $(\alpha=0)$. For all cases considered the behaviour is qualitatively similar: all curves start from unity and decrease monotonically as $\alpha$ increases. The volume phase transition  temperature (VPTT), which for PNIPAM microgels is around $T_{VPT}\sim 32\degree$C, in our model can be identified as the inflection point that develops at $\alpha \simeq 0.6$. For large values of $\alpha$, that is, when the polymer network is in a bad solvent, microgels collapse on themselves and their size plateaus at some asymptotic value. This value is known to be dependent on the crosslinker concentration as well as on the synthesis protocol~\cite{fernandez2011microgel}. Here we are able to reproduce both effects. Indeed, the change in relative size between the swollen and collapsed states is less marked as the crosslinker concentration increases or the extent of the initial confinement decreases. Therefore, the two parameters we can control, $c$ and $Z$, have the same qualitative effect on the swelling behaviour. From a quantitative standpoint, however, their effect is markedly different. Figure~\ref{fig:swelling} shows that increasing $c$ by a factor of two or three yields only a $10$ -- $20\%$ difference in the swelling behaviour, whereas a similar change in $Z$ produces microgels that are almost twice as softer.

The swelling behaviour can be characterised in more detail by looking at the internal structure of microgels as $\alpha$ increases.

\begin{figure}[h!]
\includegraphics[width=0.5\textwidth,clip]{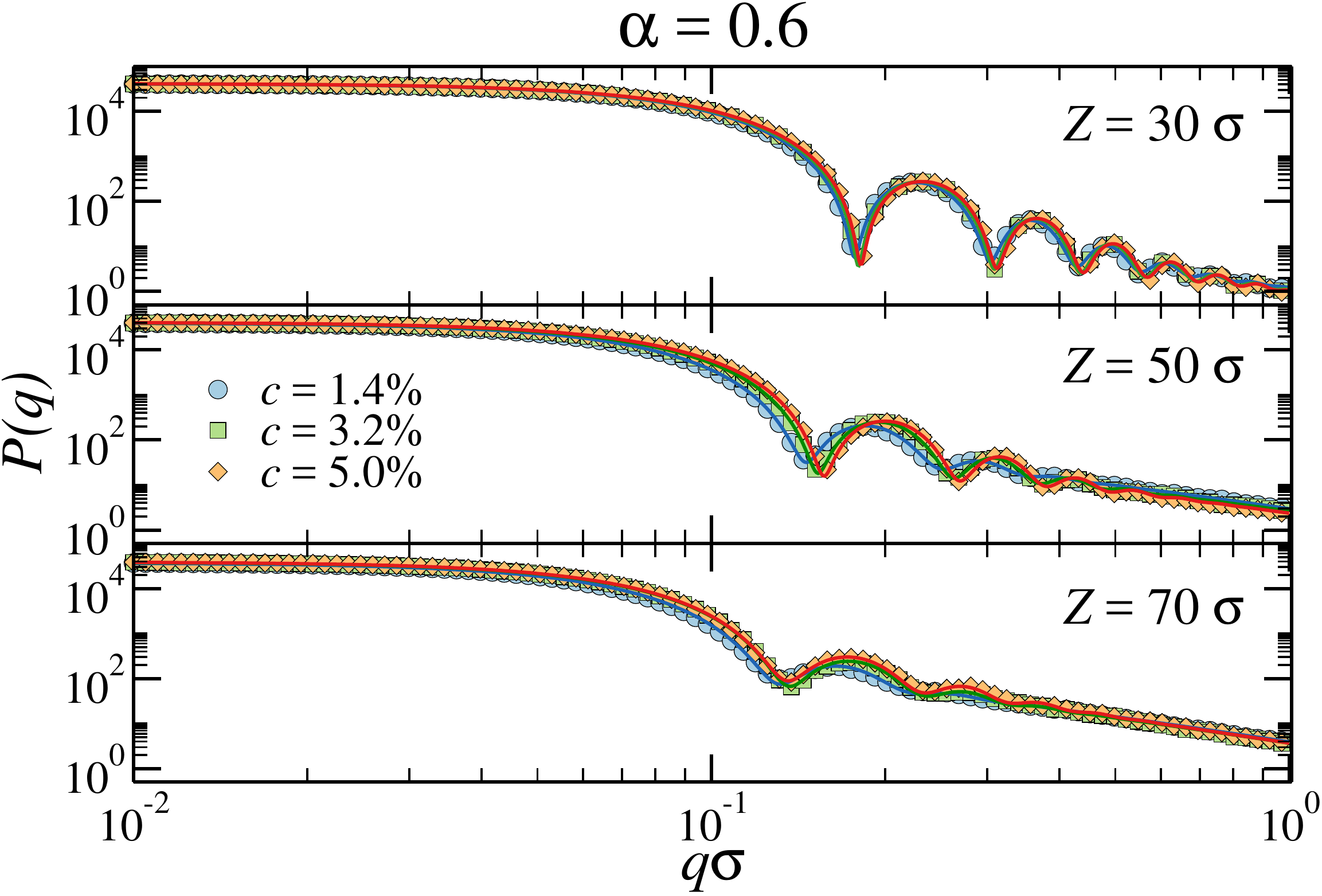}
\caption{\label{fig:pq_alpha0_6}Simulation (points) and fitted (lines) form factors $P(q)$ for microgels of different crosslinker concentrations at three different values of the radius of confinement $Z$, close to the volume phase transition ($\alpha = 0.6$).}
\end{figure}

Figure~\ref{fig:pq_alpha0_6} shows the form factors, and the accompanying fuzzy-sphere-model fits, of selected microgels for $\alpha = 0.6$, i.e. close to the VPT (see Fig~\ref{fig:swelling}). From a qualitative standpoint, the enhanced solvophobicity of the monomers do not change the trends with $c$ and $Z$ observed for the $\alpha = 0$ case, as seen in Fig.~\ref{fig:pq_alpha0}. Quantitatively, we observe a slight increase in the number and in the sharpness of the peaks, especially at small values of the initial confinement. The behaviour of $P(q)$ is again well-described by the fuzzy-sphere model for all the values of $\alpha$ considered here, making it possible to closely follow the evolution of the values of the fitting parameters through the whole volume phase transition, from the swollen to the collapsed state.

\begin{figure}[h!]
\includegraphics[width=0.5\textwidth,clip]{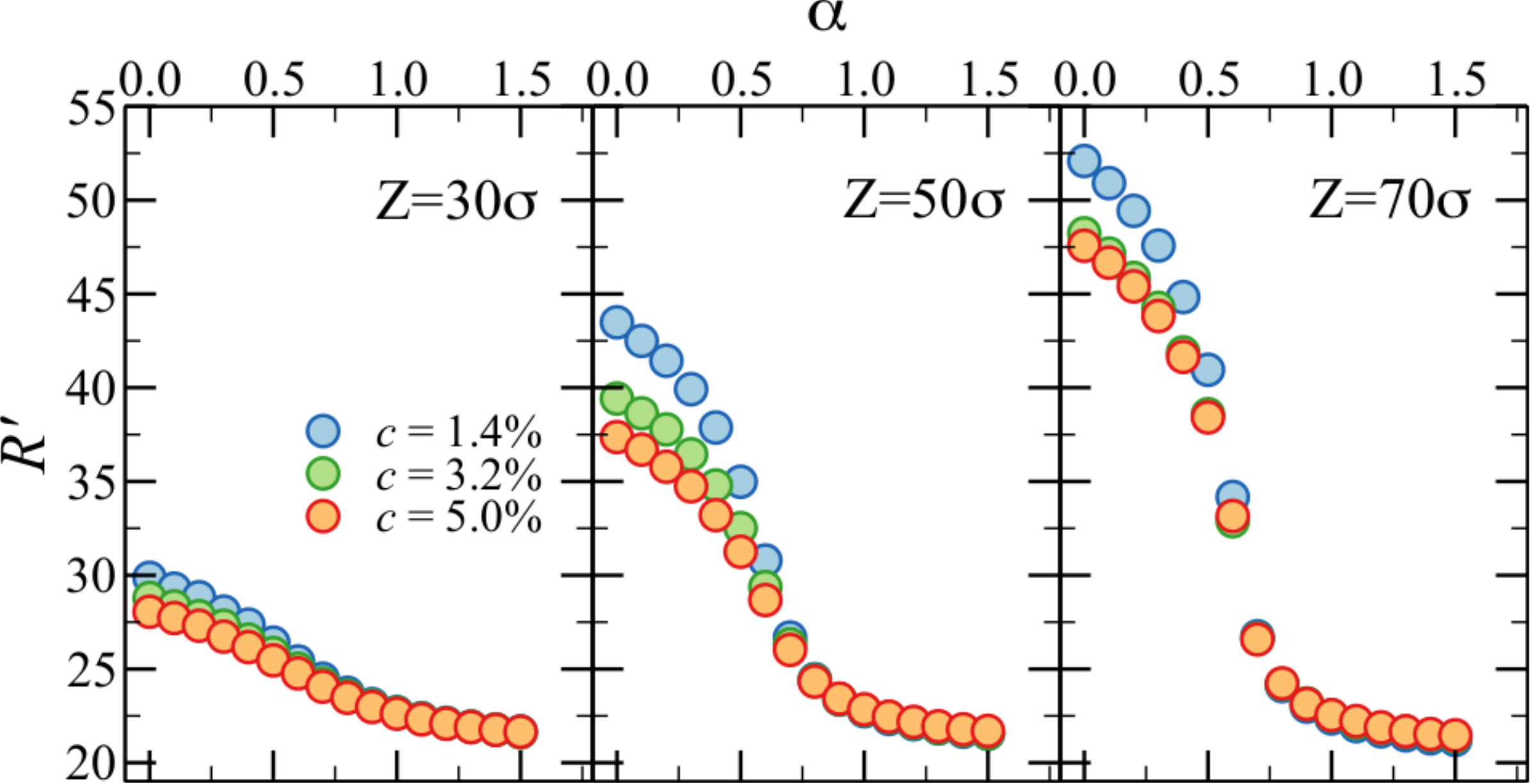}\\
\includegraphics[width=0.5\textwidth,clip]{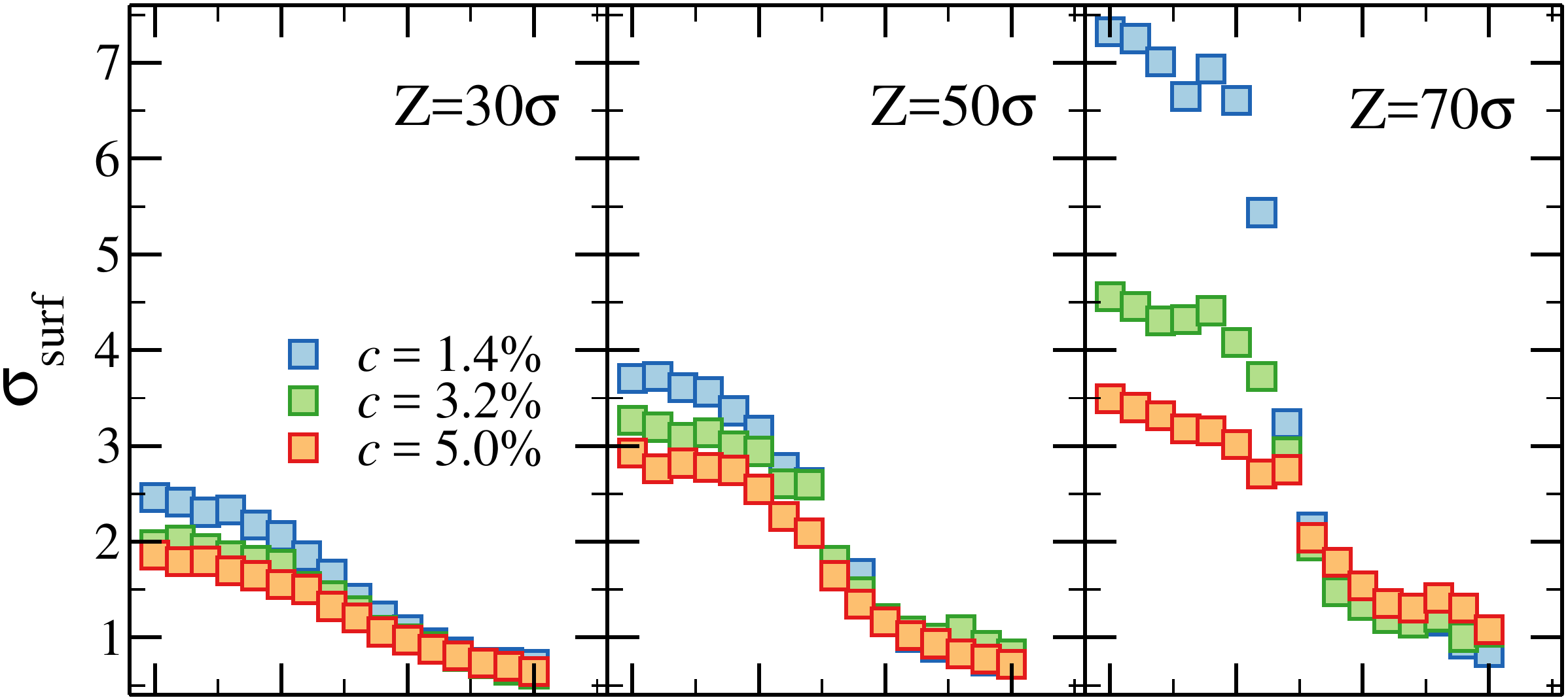}\\
\includegraphics[width=0.5\textwidth,clip]{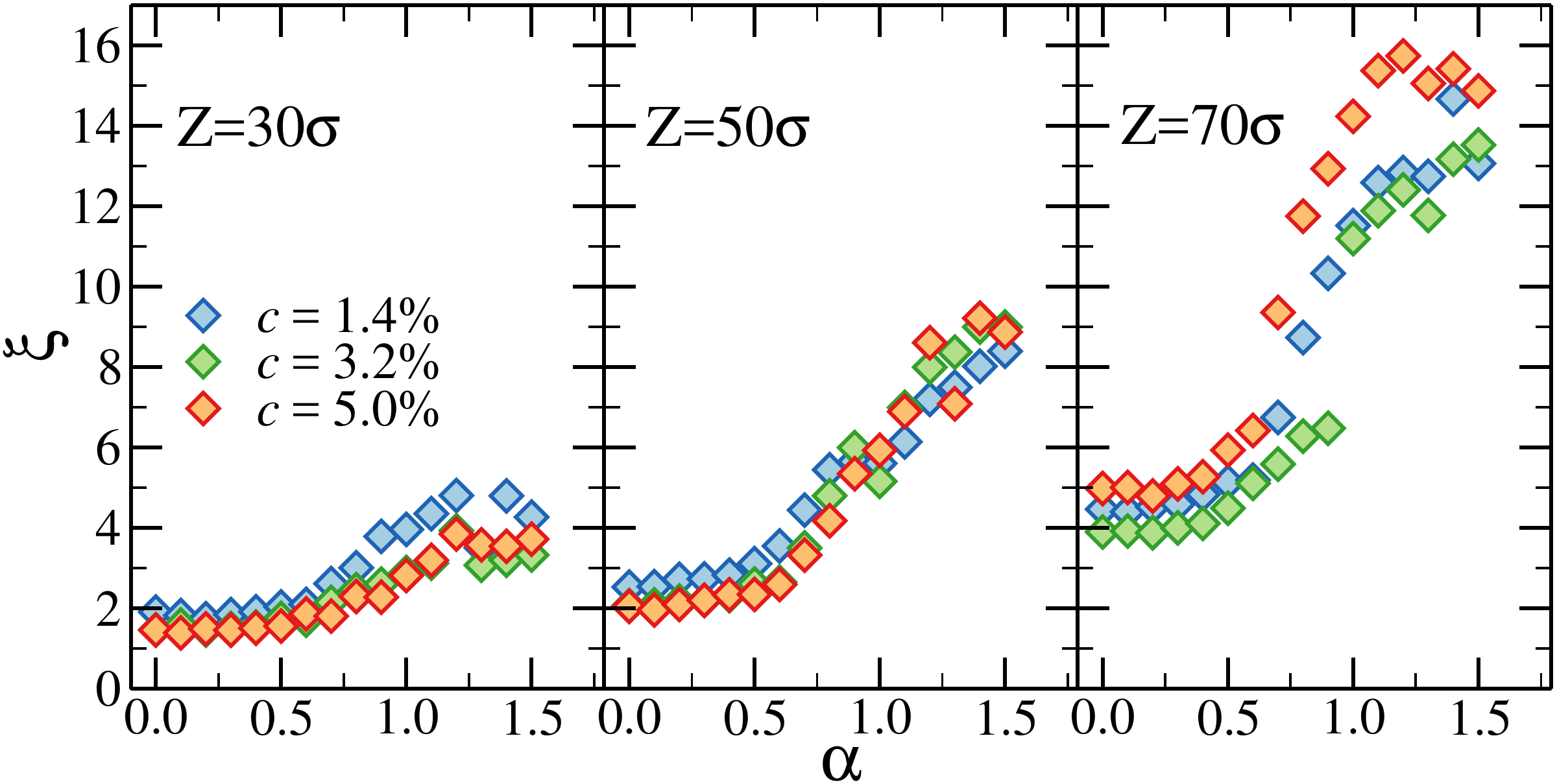}
\caption{\label{fig:fit_parameters}The fuzzy-sphere fitting parameters core radius $R'$, smearing parameter $\sigma_{\rm surf}$ and network correlation length $\xi$ of microgels generated with different crosslinker concentrations and three different values of $Z$. Lines are guides to the eye.}
\end{figure}

Figure~\ref{fig:fit_parameters} shows the core radius $R'$, the smearing parameter $\sigma_{\rm surf}$ and the network correlation length $\xi$ obtained by fitting the form factors of the microgels to the fuzzy-sphere model as the quality of the solvent varies. The effect of increasing $\alpha$ is, in all cases, monotonic: the sizes of the core and of the corona (linked to $R'$ and $\sigma_{\rm surf}$, respectively) decrease, while $\xi$ increases (within numerical noise). From a qualitative standpoint, Figure~\ref{fig:fit_parameters} shows once again that the effect of the crosslinker concentration and of the initial radius of confinement is similar: smaller $Z$ and larger $c$ tend to produce smaller (and hence less swellable) microgels and \textit{vice versa}. The behaviour of the two quantities linked to the particle size, $R'$ and $\sigma_{\rm surf}$, is qualitatively similar to the behaviour of $R_g$, as seen in Figure~\ref{fig:swelling}. However, here the contributions to the swelling given by the core and by the corona are decoupled and can be analysed separately. On one hand, the extent of the variation of $R'$ with $\alpha$ closely resembles the one of $R_g$ for all investigated values on $c$ and $Z$. On the other hand, the relative change in size of $\sigma_{\rm surf}$ due to the worsening of the solvent quality is much stronger: for the softest microgel ($c=1.4\%$, $Z = 70$), there is a sevenfold difference between the values of the smearing parameter in the swollen and collapsed states. This striking dependence on $\alpha$ originates from the low density of the corona, which is mainly composed of loosely crosslinked chains which, in the swollen state, tend to maximise the available volume by expanding, thus increasing their entropy. However, as the quality of the solvent worsens, the balance between entropy and enthalpy starts to favour the latter. As a result, the density of the outer part increases much faster than the density of the core, greatly reducing the extent of the corona, as signalled by the relative drop of $\sigma_{\rm surf}$.

The other meaningful parameter that can be extracted from the fits is the network correlation length $\xi$. It is commonly linked to the spatial heterogeneity of the polymer network over intermediate length scales. We note that $\xi$ is a phenomenological quantity which is implicitly defined by the fitting function and thus cannot be directly computed starting from the equilibrium configurations obtained in simulations. The bottom panel of Figure~\ref{fig:fit_parameters} shows that $\xi$ is a monotonically-increasing function of $\alpha$ and depends very weakly on $c$, while it is more susceptible to changes in $Z$. We note that the commonly observed behaviour of $\xi$ in experiment is to decrease with temperature, rather than increase~\cite{shibayama1992,nigro_xi}. More work is required to understand the origin of such a difference.

At very large $Z$ we observe a non-monotonic behaviour with $c$ which we ascribe to the lack of structure in these very loosely confined microgels which makes the fitting (and the interpretation of the parameters) less obvious. At high values of $\alpha$ we do not see the collapse of all the curves (regardless of $Z$ and $c$) on a single master value, as seen for the cases of $R'$ and $\sigma_{\rm surf}$. Looking at Figure~\ref{fig:pq_alpha0_6}, we see that, as the confinement gets looser and looser, form factors retain their overall shape but become less structured, displaying shallower dips and fewer oscillations. While we have not directly probed the internal structure of the collapsed microgels, we ascribe such a difference in the form factors, and hence in the values of the fitting parameter $\xi$, to an amplification of the difference in the heterogeneities of the structure which occurs during the collapse. The source of the initial difference in heterogeneity, signalled by the different values of $\xi$ at $\alpha=0$, is due to the different network structure generated during the assembly stage. Indeed, networks assembled by patchy-like particles are known to have structure factors that exhibit steeper and steeper low-$q$ behaviours as the density decreases~\cite{demichele2006dynamics,rovigatti2012structural}.

\begin{figure}[h!]
\includegraphics[width=0.5\textwidth,clip]{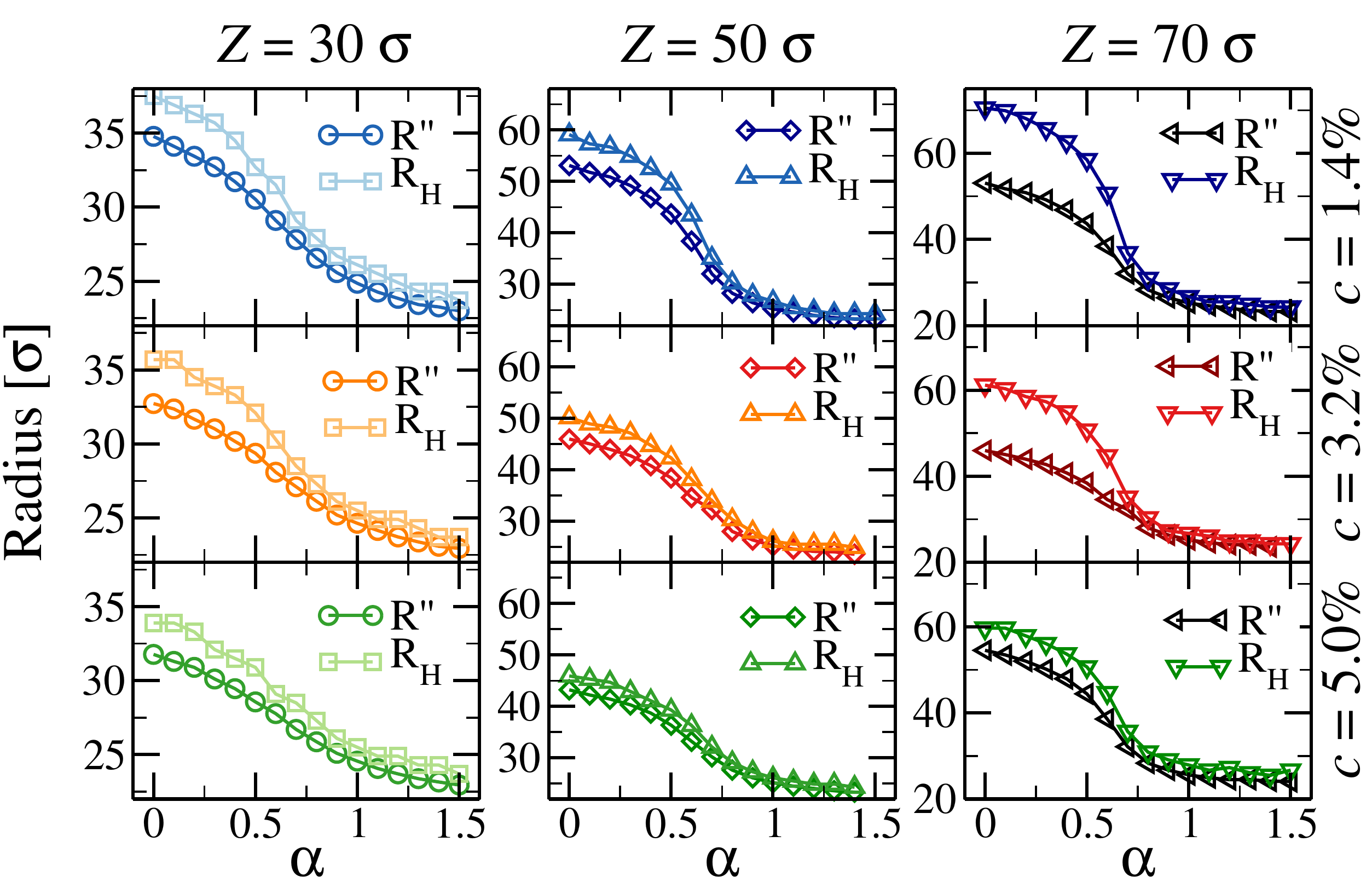}
\caption{\label{fig:RhvsRg}Fuzzy-sphere ($R''$) and hydrodynamic ($R_H$) radii as functions of $\alpha$ for microgels generated with $c = 1.4\%$, $3.2\%$ and $5\%$ and three different values of $Z$, as indicated at the right and top of the plot, respectively.}
\end{figure}

As a last point, we investigate how the total fuzzy-sphere radius $R''$ 
and the hydrodynamic radius $R_H$, which is defined as the radius at which $\rho(r)/\rho(0) = 10^{-3}$, as done in Ref.~\cite{gnan2017silico}, change with $c$, $Z$ and $\alpha$. We use $R''$ and $R_H$ as proxies for the SANS and DLS radii measured experimentally, respectively. These quantities are known to differ, with the latter being always larger than the former~\cite{fernandez2011microgel}. Figure~\ref{fig:RhvsRg} shows $R''$ and $R_H$ as functions of the solvent quality for microgels generated with different $c$ and $Z$. Both quantities monotonically decrease with $\alpha$ and $c$ and monotonically increase with $Z$. Interestingly, the hydrodynamic radius exhibits a larger variation across the volume phase transition. In a good solvent, \textit{i.e.} $\alpha \to 0$, dangling ends or very long closed loops can stick out of the corona and increase the effective size of particle, thereby boosting the value of $R_H$. When the presence of these loosely connected chain portions is higher, as it is the case for larger values of $Z$ and smaller values of $c$, the effect is enhanced. However, in the collapsed state, the dangling ends are mostly attached to the particle surface and thus do not contribute significantly to the value of $R_H$. As a result, $R_H$ approaches $R''$ as $\alpha$ increases.

\section{Conclusions}
In this work we have investigated the internal structure of numerical microgels across the volume phase transition as the crosslinker concentration $c$ and the initial assembly conditions vary. The flexible protocol we have recently introduced to generate these microgels is based on the self-assembly of a confined mixture of bivalent and tetravalent patchy monomers that gives rise to an almost fully bonded network assembled in a spherical confinement ($95\%-99\%$ of particles belong to the largest cluster).

For the present study, we design microgels with a total number of monomers $N\sim 41000$; by assuming each coarse-grained monomer to have a size comparable with the Kuhn length, the resulting microgels are the numerical analogue of experimental PNIPAM microgels of small size. A direct comparison with experimental data, made possible by the small size of the employed microgels, is discussed in a recent work~\cite{gnan2017silico}. We plan to draw similar comparisons with other small microgels of different crosslinker concentrations in the near future. We have investigated three crosslinker concentrations, namely $c=1.4\%, 3.2\%, 5.0\%$, and several initial confining radii $Z$, ranging from $Z=30\sigma$ to $Z=70\sigma$. We find that the internal architecture of microgel particles is controlled qualitatively in the same way  by both $c$ and $Z$; large $c$ and small $Z$ values generate compact microgels, while large confining radii and small $c$ values enhance the formation of more heterogeneous and softer networks. However, while $Z$ acts on the degree of entanglement of the polymer chains in the network, the crosslinker concentration modify the chain-length distribution. We have shown that the latter can be derived theoretically from the Flory theory in which we have included the requirement that bonds among crosslinkers are absent. The effect of $Z$ and $c$ on the microgel structure is also reflected in the numerical form factors: we find that strongly confined microgels or microgels with high $c$ display form factors with several sharp peaks, an indication of the presence of a  structured crosslinked network. On the other hand, small crosslinker concentration and large confining radii result in form factors with few shallow peaks. This is valid also for larger values of the solvophobic parameter $\alpha$, which plays the role of a temperature and controls the degree of swelling of the microgel particles. By fitting the form factors to the fuzzy-sphere model we have extracted the density profiles of the microgels, finding that they are  in good agreement with the numerical density profiles directly calculated from simulations. Interestingly, we observe from such fits that, in the collapsed state beyond the VPT, a difference between  microgels generated at large and small $Z$  still persists, even if the radius of gyration is almost the same for all $Z$  and $c$ values.
This difference could be related to non-equilibrium effects occurring in the monomer dynamics at high values of $\alpha$; indeed, during the collapse monomers could get trapped in a disordered arrested state whose microscopic structure would depend on the initial conditions of the microgel structure, and hence on $Z$. Further investigation on the monomer dynamics is needed to better understand this point.

Finally, we have also investigated the difference between the total radius of the fuzzy sphere and its hydrodynamic radius. The latter accounts for the presence of dangling ends: a large difference between $R_H$ and $R''$ indicates the presence of several of these loosely connected portions of the outer corona. We have found that $R_H$ and $R''$ differ more for large confining radii and small crosslinker concentration. However, in the collapsed state the long chains, that were free to move  for small $\alpha$ outside of the corona radius, collapse on themselves and attach onto the core of the microgel, thus almost cancelling the difference between  $R_H$ and $R''$.
The analysis proposed in this work allows to better understand the structural properties of microgels generated with our novel synthesis protocol in an effort towards the design of realistic microgels, which encodes the correct properties of experimental microgel particles. This effort is ultimately aimed at extracting reliable effective interactions and to go beyond the widely used, but over-simplified, Hertzian model~\cite{mohanty2014effective}.

\section*{Acknowledgements}
LR and NG contributed equally to this work. We acknowledge support from the European Research Council (ERC Consolidator Grant 681597, MIMIC). 

\bibliography{library}

\end{document}